\documentclass[aps,preprint,amsmath,amssymb]{revtex4}

\usepackage{graphicx}% Include figure files
\usepackage{dcolumn}% Align table columns on decimal point
\usepackage{bm}% bold math

\begin{document}
\title{\Large \bf Polarizations of two vector mesons in $B$ decays}
% $B\to (\rho, \omega, \phi) K^*$ and $B\to \rho (\omega) \rho (\omega)$ decays}
\date{\today}
\author{\large \bf
Chuan-Hung~Chen$^{1,2}$\footnote{Email: phychen@mail.ncku.edu.tw} }
\affiliation{
$^{1}$Department of Physics, National Cheng-Kung University, Tainan, 701 Taiwan \\
$^{2}$ National Center for Theoretical Sciences, Taiwan }

\begin{abstract}
Inspired by the small longitudinal polarizations (LPs) of $B\to K^*
\phi$ decays observed by BELLE and BABAR, we revise the theoretical
uncertainties of perturbative QCD approach for determining hard
scales of $B$ decays, we find that the LPs of $B\to K^* \phi$ could
approach to $60\%$ while the branching ratios (BRs) could be around
$9\times 10^{-6}$. In addition, we also study the BRs and
polarization fractions of $B\to \rho (\omega) \rho (\omega)$ and
$B\to \rho (\omega) K^*$ decays. For those tree dominant and
color-allowed processes in $B\to \rho (\omega) \rho (\omega)$
decays, we get that the BRs of $(\rho^{+} \rho^{-},\, \rho^{0}
\rho^{+},\, \omega \rho^{+})$ are $(23.06,\, 11.99,\, 14.78)\times
10^{-6}$ while their LPs are close to unity. Interestingly, due to
significant tree contributions, we find that the BR(LP) of $\rho^{-}
K^{*+}$ could be around $10.13 \times 10^{-6}(60\%)$; and due to the
tree and electroweak penguin, the BR(LP) of $\omega K^{*+}$ could be
around $5.67 \times 10^{-6}(61\%)$.

\end{abstract}
\maketitle

Since the transverse polarizations (TPs) of vector mesons are
associated with their masses, by naive estimations, we can easily
obtain that the longitudinal polarization (LP) of the two light
vector mesons produced by $B$ decay is approaching to unity. The
expectation is confirmed by BELLE \cite{belle1} and BABAR
\cite{babar1-1, babar1-2} in $B\to \rho(\omega) \rho$ decays, in
which the longitudinal parts occupy over $88\%$. Furthermore,
TP(LP) could be large(small) while the final states include heavy
vector mesons. The conjecture is verified in $B\to J/\Psi K^*$
decays \cite{belle2,babar2}, in which the longitudinal
contribution is only about $ 60\%$. However, the rule for small LP
seems to be broken in  $B\to \phi K^*$ decays. From the recent
measurements of BELLE \cite{belle3} and BABAR
\cite{babar1-1,babar3}, summarized in the Table \ref{tab:pol}, it
is quite clear that the LPs of $B\to K^* \phi$ are only around
$50\%$. According to the observations, many mechanisms are
proposed to solve the puzzle, where the methods include not only
new QCD effects \cite{QCD} but also the effects of the extension
of the standard model (SM) \cite{newphys,CG_PRD71}.
%%%%%%%%%%%%%%%%%%%%%%%%%%
%%%%%%%%%%%%%%%%%%%%%%%%%%%%%%%%%%%%%%%%%
\begin{table}[hptb]
\caption{\label{tab:pol} The branching ratios (in units of
$10^{-6}$), polarization fractions and relative phases for $B\to
\phi K^*$. }
\begin{ruledtabular}
\begin{tabular}{cccc}
Mode & Observation & BELLE & BABAR \\ \hline $K^{*+}\phi$ & BR &
$10.0^{+1.6+0.7}_{-1.5-0.8}$ & $12.7^{ + 2.2}_{ - 2.0}\pm 1.1$ \\
$$& $R_{L}$ & $0.52 \pm 0.08 \pm 0.03$ & $0.46 \pm 0.12 \pm 0.03$\\
$$ & $R_{\perp}$ & $0.19\pm 0.08\pm 0.02$ &
$$ \\
$$ & $\phi_{\parallel}(rad)$ & $2.10\pm 0.28\pm 0.04$ &
$$ \\
$$ & $\phi_{\perp}(rad)$ & $2.31\pm 0.20 \pm 0.07$ &
$$ \\
\hline $K^{*0} \phi$&  BR & $6.7^{+2.1+0.7}_{-1.9-1.0}
$ & $9.2\pm 0.9 \pm 0.5$ \\
$$&$R_{L}$ & $0.45 \pm 0.05 \pm 0.02$ & $0.52 \pm 0.05 \pm 0.02$\\
$$ & $R_{\perp}$ & $0.30 \pm 0.06 \pm 0.02$ & $0.22\pm 0.05 \pm
0.02$\\
 $$ & $\phi_{\parallel}(rad)$ & $2.39 \pm 0.24 \pm 0.04$ &
$2.34^{+0.23}_{-0.20}\pm 0.05$ \\
$$ & $\phi_{\perp}(rad)$ & $2.51 \pm 0.23 \pm 0.04$ & $2.47\pm 0.25 \pm 0.05$
 \\
\end{tabular}
\end{ruledtabular}
\end{table}
%%%%%%%%%%%%%%%%%%%%%%%%%%%%%%%%%%%%%%%%%%%%%%%%%%%%%%%%%

It is known that most proposals to solve the anomalous polarizations
only concentrate on how to make the LPs of $B\to K^* \phi$ be small.
It is few to analyze the problem by combing other decays such as the
decays $B\to \rho (\omega) \rho (\omega)$ and $B\to \rho (\omega)
K^*$ etc. That is, maybe we can invent a way to solve the anomalies
in $K^* \phi$, however, we still don't have the definite reason to
say why the considering effects cannot contribute to $\rho(\omega)
\rho(\omega)$ or $\rho(\omega) K^*$ significantly. By this
viewpoint, in this paper, we are going to reanalyze the decays $B\to
K^* \phi$ in terms of perturbative QCD(PQCD) \cite{PQCD1,PQCD2}
approach in the SM. By revising the theoretical uncertainties of
PQCD, which come from the man-made chosen conditions for hard scales
of $B$ decays, we will show how well we can predict and how close we
can reach in theoretical calculations, while the processes of light
mesons production are assumed to be dominated by the short-distant
effects. We note that the wave functions of mesons, representing the
nonpertubative QCD effects, are assumed to be known and obtained by
the QCD sum rules \cite{DA1,DA2}. Moreover, according to the
improving conditions, we also make the predictions on the decays
$B\to \rho (\omega) \rho(\omega)$ and $B\to \rho (\omega) K^*$.

\begin{figure}[htbp]
\includegraphics*[width=3.in]{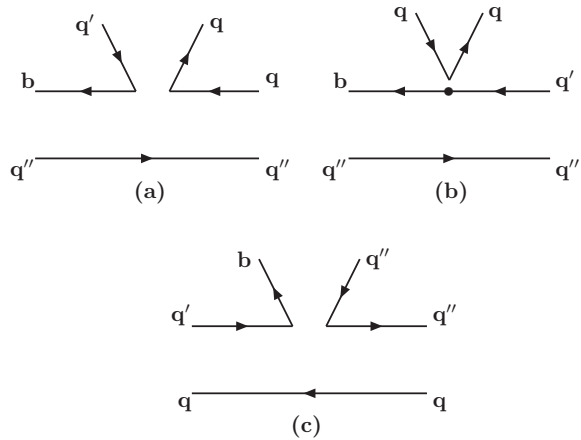}  \caption{For $b\to q^{\prime} q \bar{q}$ decays, the flavor diagrams
(a) and (b) stand for the emission topologies while (c) is
annihilation topology.  }
 \label{fig:flavor}
\end{figure}

Although the effective interactions, governing the transition decays
$b\to s (d)$ at the quark level, are well known, to be more clear
for explanation, we still write them out to be \cite{BBL}
\begin{equation}
H_{{\rm eff}}={\frac{G_{F}}{\sqrt{2}}}\sum_{q=u,c}V_{q}\left[
C_{1}(\mu) O_{1}^{(q)}(\mu )+C_{2}(\mu )O_{2}^{(q)}(\mu
)+\sum_{i=3}^{10}C_{i}(\mu) O_{i}(\mu )\right] \;,
\label{eq:hamiltonian}
\end{equation}
where $V_{q}=V_{qq^{\prime}}^{*}V_{qb}$ are the
Cabibbo-Kobayashi-Maskawa (CKM) \cite{CKM} matrix elements, the
subscript $q^{\prime}$ could be $s$ or $d$ quark and the operators
$O_{1}$-$O_{10}$ are defined as
\begin{eqnarray}
&&O_{1}^{(q)}=(\bar{q}^{\prime}_{\alpha}q_{\beta})_{V-A}(\bar{q}_{\beta}b_{\alpha})_{V-A}\;,\;\;\;\;\;
\;\;\;O_{2}^{(q)}=(\bar{q}^{\prime}_{\alpha}q_{\alpha})_{V-A}(\bar{q}_{\beta}b_{\beta})_{V-A}\;,
\nonumber \\
&&O_{3}=(\bar{q}^{\prime}_{\alpha}b_{\alpha})_{V-A}\sum_{q}(\bar{q}_{\beta}q_{\beta})_{V-A}\;,\;\;\;
\;O_{4}=(\bar{q}^{\prime}_{\alpha}b_{\beta})_{V-A}\sum_{q}(\bar{q}_{\beta}q_{\alpha})_{V-A}\;,
\nonumber \\
&&O_{5}=(\bar{q}^{\prime}_{\alpha}b_{\alpha})_{V-A}\sum_{q}(\bar{q}_{\beta}q_{\beta})_{V+A}\;,\;\;\;
\;O_{6}=(\bar{q}^{\prime}_{\alpha}b_{\beta})_{V-A}\sum_{q}(\bar{q}_{\beta}q_{\alpha})_{V+A}\;,
\nonumber \\
&&O_{7}=\frac{3}{2}(\bar{q}^{\prime}_{\alpha}b_{\alpha})_{V-A}\sum_{q}e_{q} (\bar{q}%
_{\beta}q_{\beta})_{V+A}\;,\;\;O_{8}=\frac{3}{2}(\bar{q}^{\prime}_{\alpha}b_{\beta})_{V-A}
\sum_{q}e_{q}(\bar{q}_{\beta}q_{\alpha})_{V+A}\;,  \nonumber \\
&&O_{9}=\frac{3}{2}(\bar{q}^{\prime}_{\alpha}b_{\alpha})_{V-A}\sum_{q}e_{q} (\bar{q}%
_{\beta}q_{\beta})_{V-A}\;,\;\;O_{10}=\frac{3}{2}(\bar{q}^{\prime}_{\alpha}b_{\beta})_{V-A}
\sum_{q}e_{q}(\bar{q}_{\beta}q_{\alpha})_{V-A}\;, \label{eq:ops}
\end{eqnarray}
with $\alpha$ and $\beta$ being the color indices. In Eq.
(\ref{eq:hamiltonian}), $O_{1}$-$O_{2}$ are from the tree level of
weak interactions, $O_{3}$-$O_{6}$ are the so-called gluon penguin
operators and $O_{7}$-$O_{10}$ are the electroweak penguin
operators, while $C_{1}$-$C_{10}$ are the corresponding WCs. Using
the unitarity condition, the CKM matrix elements for the penguin
operators $O_{3}$-$O_{10}$ can also be expressed as
$V_{u}+V_{c}=-V_{t}$. To describe the decay amplitudes for $B$
decays, we have to know  not only the relevant effective weak
interactions but also all possible topologies for the specific
process. In terms of penguin operators, we display the general
involving flavor diagrams for $b\to q^{\prime} q \bar{q}$ in Fig.
\ref{fig:flavor}, where (a) and (b) denote the emission topologies
while (c) is the annihilation topology. The flavor $q$ in
Fig.\ref{fig:flavor}(a) and (b) is produced by gauge bosons and
could be $u$, or $d$ or $s$ quark if the final states are the light
mesons; however, $q^{\prime \prime}$ stands for the spectator quark
and could only be $u$ or $d$ quark, depending the $B$ meson being
charged or neutral one.
%%%%%%%%%%%%%%%%%%%%%%%%%%%%%%%%%%%%%%%%%%%%%%%%%%%%%%%%%%%%%%%%%%%%%%%%%
%%%%%%%%%%%%%%%%%%%%%%%%%%%%%%%%%%%%%%%%%%%%%%%%%%%%%%%%%%%%%%%%%%%%%%%%%
However, the role of $q$ and $q^{\prime \prime}$ in
Fig.~\ref{fig:flavor}(c) is reversed so that $q=u$, or $d$, or $s$
is the spectator quark while $q^{\prime \prime}=u$ or $d$ is
dictated by gauge interactions. Since the matrix elements obtained
by the Fierz transformation of $O_{3,4}$ are the same as those of
$O_{1,2}$, we don't further consider the flavor diagrams for tree
contributions.

In the beginning, we first pay attention to $B\to K^* \phi$ decays.
Although there are charged and neutral modes in $B\to K^* \phi$
decays, because the differences in charged and neutral modes are
only the parts of small tree annihilation, for simplicity our
discussions will concentrate on the decay $B_{d}\to K^{*0} \phi$. As
known that at quark level, the decay corresponds to $b\to s s
\bar{s}$; thus, by the flavor diagrams, we have $q=q^{\prime}=s$ and
$q^{\prime \prime}=d$. According to our previous
results~\cite{CKLPRD66}, the helicity amplitude could be expressed
by
\begin{eqnarray}
{\cal M}^{(h)}
% \epsilon_{1\mu}^{*}(h)\epsilon_{2\nu}^{*}(h)
%\left[ a \,g^{\mu\nu} + {b \over M_{V_1} M_{V_2}} P_2^\mu P_1^\nu +
%i{c \over M_{V_1} M_{V_2} } \epsilon^{\mu\nu\alpha\beta} P_{1\alpha}
%P_{2\beta}\right]\;
%\nonumber \\
\equiv m_{B}^{2}{\cal M}_{L} + m_{B}^{2}{\cal M}_{N}
\epsilon^{*}_{1}(t)\cdot\epsilon^{*}_{2}(t) +i{\cal
M}_{T}\epsilon^{\alpha \beta\gamma \rho}
\epsilon^{*}_{1\alpha}(t)\epsilon^{*}_{2\beta}(t) P_{1\gamma
}P_{2\rho } \label{eq:hamp}
\end{eqnarray}
with the convention $\epsilon^{0123} = 1$, where the superscript $h$
is the helicity, ${\cal M}_{h}$ is the amplitude with helicity $h$
and it's explicit expression could be found in Ref.~\cite{CKLPRD66},
the subscript $L$ stands for $h=0$ component while $N$ and $T$
express another two $h=\pm 1$ components, $P_{1(2)}$ denote the four
momenta of vector mesons, and $\epsilon^*_{1}(t)\cdot
\epsilon^{*}_{2}(t)=1$ with $t=\pm 1$. Hence, each helicity
amplitude could be written as \cite{CKLPRD66}
\begin{eqnarray}
H_{0}&=&m^{2}_{B} {\cal M}_L\;,
\nonumber\\
H_{\pm}&=& m^{2}_{B} {\cal M}_{N} \mp  m_{V_1} m_{V_2}
\sqrt{r^2-1}{\cal M}_{T}\;,
\end{eqnarray}
and $r=P_{1}\cdot P_{2}/(m_{V_1}m_{V_2})$ in which $m_{V_{1(2)}}$
are the masses of vector mesons. Moreover, we can also write the
amplitudes in terms of polarizations as
\begin{eqnarray}
A_{L}=H_{0} \ \ \ A_{\parallel(\perp)}=\frac{1}{\sqrt{2}}(H_{-} \pm
H_{+}). \label{pol-amp}
\end{eqnarray}
The relative phases are defined as $\phi_{\parallel(\perp)}=Arg(A_{\parallel(\perp)}/A_{0})$.
Accordingly, the polarization fractions (PFs) can be defined as %
 \begin{eqnarray}
R_i=\frac{|A_i|^2}{|A_L|^2+|A_{\parallel}|^2+|A_{\perp}^2|}\,\ \
(i=L,\parallel,\perp)\,. \label{eq:pol}
 \end{eqnarray} %
Since we have derived the formalisms for the decay amplitudes ${\cal
M}_{L,N,T}$ by PQCD approach in Ref.~\cite{CKLPRD66}, in our
following discussions, we only concentrate on the theoretical
uncertainties of PQCD.

It is known that by PQCD the transition amplitude is factorized into
the convolution of hadron wave functions and the hard amplitude of
the valence quarks, in which the wave functions absorb the infrared
divergences and represent the effects of nonperturbative QCD. With
including the transverse momentum of valence quark, $k_{T}$, the
factorization formula for
the decay of $B$ meson could be briefly described as \cite{PQCD2}%%%
\begin{eqnarray}
H_{r}(m_{W},\mu)H(t,\mu)\Phi(x,P,b,\mu)&=&c(t) H(t,t)
\Phi(x,b,1/b)\nonumber \\
&&\times
\exp\left[-s(P,b)-\int^{t}_{1/b}\frac{d\bar{\mu}}{\bar{\mu}}
\gamma_{\Phi}(\alpha_{s}(\bar\mu)) \right]
\end{eqnarray}
where $H_{r}(m_W,\mu)$ and $H{(t,\mu)}$ denote the renormalized hard
parts which the running scale starts from $m_{W}$ and typical hard
scale $t$, respectively, $\Phi(x,P,b,\mu)$ is the wave function of
meson, $c(t)$ is the effective Wilson coefficient, $b$ is the
conjugate variable of $k_{T}$, $s(P,b)$ is Sudakov factor for
suppressing the radiative corrections at large $b$ parameter, and
$\gamma_{\Phi}$ stands for the anomalous dimension of valence quark.
Clearly, for calculating the decay amplitudes of $B$ decays, we have
to determine the typical scale which dictates the decaying scale of
$B$ meson. To illustrate the chosen hard scale in conventional PQCD,
we take the transition matrix element $\langle M(P_{2})| \bar{b}
\gamma_{\mu} q |B(P_{1})\rangle$ as the example. As usual,
the condition for the hard scale is set to be %
\begin{eqnarray}
t={\rm max}\left(\sqrt{x_{1}m^2_{B}}, \sqrt{x_2
m^2_{B}},1/b_1,1/b_2\right), \label{eq:scale}
\end{eqnarray}
where $x_{1(2)}$ are the momentum fraction carried by the quark of
$B$(M) meson. Since the allowed range of momentum fraction is
between 0 and 1, therefore the value of hard scale could be less
than $1$ GeV. However, the wave functions such as twist-2 wave
function expressed by
\begin{eqnarray}
 \Phi \left( {x,\mu ^2 } \right) = 6x\left( {1 - x} \right)\left(
{1 + \sum\limits_{n = 1}^\infty  {a_n \left( {\mu ^2 } \right){\rm{
}}} C_n^{3/2} \left( {2x - 1} \right)} \right),
\end{eqnarray}
are expanded by the Gegenbauer polynomials; and the scale-dependent
coefficients are usually estimated at $\mu=1$ GeV. That is, the
physics below $1$ GeV belongs to nonperturbative region and hard
scale should end up at this scale. Consequently, we regard that the
condition of Eq.~(\ref{eq:scale}) should be revised to be
\begin{eqnarray}
t={\rm max}\left(\sqrt{x_{1}m^2_{B}}, \sqrt{x_2
m^2_{B}},1/b_1,1/b_2,\bar \Lambda\right) \label{eq:hardscale}
\end{eqnarray}
where $\bar\Lambda$ indicates the cutoff for distinguishing the
region of perturbation and nonperturbation, i.e. below $\bar\Lambda$
the physics is dominated by nonperturbative effects. Roughly, the
order of magnitude of the hard scale could be estimated by the
momentum of exchanged hard gluon as $t \sim \sqrt{x_1 x_2
m^{2}_{B}}$. It is known that $x_{1}\sim (m_{B}-m_{b})/m_{B}$ and
$x_{2}\sim O(1)$. By taking $x_{1}=0.16$, $x_{2}=0.5$ and
$m_{B}=5.28$ GeV, the average value of hard scale could be estimated
to be around $\bar t \sim 1.5$ GeV. Besides the chosen condition for
hard scale and wave functions of light mesons, the remaining
uncertainties of PQCD are the shape parameter $\omega_{B}$ of the
$B$ meson wave function and the parametrization of threshold
resummation, denoted by $S_{t}(x)=2^{1+2c}\Gamma(1+2
c)[x(1-x)]^{c}/\sqrt{\pi}(1+c)$ \cite{CKLPRD66}. In our following
numerical estimations, we will set $\omega_{B}=c=0.4$. Hence,
according to the wave functions derived by QCD sum rules \cite{DA1}
and using $f^{(T)}_{K^*}=210 (170)$ MeV, the values of $B\to K^*$
form factors, defined by \cite{CG_NPB}
\begin{eqnarray}
\langle M(P_{2},\epsilon )| \bar{b}\gamma_{\mu }q| B%
(P_{1})\rangle &=&i\frac{V(q^{2})}{m_{B}+m_{M}}\varepsilon _{\mu
\alpha \beta \rho }\epsilon ^{*\alpha }P^{\beta }q^{\rho },  \nonumber \\
\langle M(P_{2},\epsilon )| \bar{b}\gamma_{\mu } \gamma_{5} q| B
(P_{1})\rangle &=&2m_{M}A_{0}(q^{2})\frac{\epsilon ^{*}\cdot q}{%
q^{2}}q_{\mu }+( m_{B}+m_{M}) A_{1}(q^{2})\Big( \epsilon
_{\mu }^{*}-\frac{\epsilon ^{*}\cdot q}{q^{2}}q_{\mu }\Big)  \nonumber \\
&&-A_{2}(q^{2})\frac{\epsilon ^{*}\cdot q}{m_{B}+m_{M}}\Big( P_{\mu }-%
\frac{P\cdot q}{q^{2}}q_{\mu }\Big), \label{eq:formfactors}
\end{eqnarray}
are given in Table~\ref{tab:bksformfactor}, where $M$ and $m_{M}$
denote the vector meson and it's mass, $P=P_{1}+P_{2}$ and
$q=P_{1}-P_{2}$. In the table, for comparison, we also show the
results of quark model (QM) \cite{MS}, light-cone sum rules (LCSR)
\cite{DA2}, and light-front quark model (LFQM) \cite{LF}.
%%%%%%%%%%%%%%%%%%%%%%%%%%%%%%%%%%%%%%%%%%%%%%%%%%%%%%%%%%%%%%%%%
\begin{table}[hptb]
\caption{ Form factors for $B\to K^*$ at $q^2=0$ in various QCD
models. }\label{tab:bksformfactor}
\begin{ruledtabular}
\begin{tabular}{ccccc}
 Model & $V( 0)$ & $A_{0}( 0) $ & $A_{1}(0) $ & $A_{2}( 0) $
 \\ \hline
 QM \cite{MS} & $0.44$ & $0.45$ & $0.36$ & $0.32$  \\ \hline
  LCSR \cite{DA2} & $0.41$ & $0.37$ & $0.29$ & $0.26$ \\ \hline
  LFQM \cite{LF} &  $0.31$ &$0.31$ & $0.26$ &$0.24$ \\ \hline
 PQCD \cite{CG_NPB} & $0.34$ &$0.37$ & $0.23$ &$0.22$  \\
 \end{tabular}
\end{ruledtabular}
\end{table}
In terms of the formulas, which are derived in Ref.~\cite{CKLPRD66}
and have included nonfactorizable and annihilation effects, and by
taking $V^{*}_{us}V_{ub}=A\lambda^{3} R_{b}e^{-i\phi_{3}}$ and
$V_{tb}V^{*}_{ts}=-A\lambda^{2}$ with $A=0.82$, $\lambda=0.224$,
$R_{b}=0.38$ and $\phi_{3}=63^{\circ}$, the calculated BR, PFs, and
$\phi_{\parallel(\perp)}$ of $B_{d}\to K^{*0} \phi$ with different
values of $\bar\Lambda$ are presented in Table~\ref{tab:br-po}.
Although there exist other chosen conditions for nonfactorized and
annihilated parts, since the conditions are similar to
Eq.~(\ref{eq:scale}), we neglected showing them. The details could
be referred to Ref.~\cite{CKLPRD66}.
%%%%%%%%%%%%%%%%%%%%%%%%%%%%%%%%%%%%%%%%%%%%%%%%%%%%%%%%%%%%%%%%%
\begin{table}[hptb]
\caption{\label{tab:br-po} BR (in units of $10^{-6}$), PFs and
relative phases of $B_{d}\to K^{*0} \phi$ for $\bar\Lambda=0,\,
1.0,\, 1.3$ and $1.6$ GeV. }
\begin{ruledtabular}
\begin{tabular}{ccccccc}
$\bar \Lambda$  & $BR$& $R_{L}$ & $R_{\parallel}$ & $R_{\perp} $ &
$\phi_{\parallel}(rad)$ &  $\phi_{\perp}(rad)$
 \\
\hline
    0 & $14.54$ & $0.71$ & $0.16$ & $0.13$& $ 2.48$ & $2.47$
    \\
     1.0 & $10.32$ & $0.65$ & $0.19$ & $0.16$& $2.33$ & $2.32$ \\
     1.3 & $8.91$ & $0.63$ & $0.20$ & $0.17$& $2.27 $ & $2.26$ \\
     1.6 & $7.69$ & $0.61$ & $0.21$ & $0.18$& $2.22$ & $2.21$ \\
\end{tabular}
\end{ruledtabular}
\end{table}
%%%%%%%%%%%%%%%%%%%%%%%%%%%%%%%%%%%%%%%%%%%%%%%%%%%%%%%%%%%%%%%%%
In the table, we have set $\bar\Lambda=0$ as the old chosen
conditions for the hard scales. From the table, we clearly see that
the BR and $R_{L}$ are decreasing while $\bar\Lambda$ is increasing.
If we regard $\bar t \sim \bar\Lambda\sim 1.5$ GeV, we obtain that
the $R_{L}$ of $B\to K^{*0} \phi$ could be around $62\%$ while the
BR could be $8\times 10^{-6}$. Since the errors of neutral $B$ decay
are still big, if we use the observed world averages of charged
mode, which they are $BR=(9.7\pm1.5)\times 10^{-6}$ and $R_L=0.50\pm
0.07$ \cite{HFAG}, as the illustration, we find that our $R_{L}$ has
approached to the upper bound of world average of $B_{u}\to
K^{*+}\phi$ while the BR is close to the lower bound. Clearly, by
using Eq.~(\ref{eq:hardscale}), we can improve our results to be
more close to the indications of data. Furthermore, in order to
understand the influence of nonfactorizable and annihilation
effects, we present the results without either and both
contributions in Table~\ref{tab:nonfa}. By the results, we could see
nonfactorizable and annihilation contributions play important role
on the PFs, especially, the annihilation effects. The brief reason
is that the penguin dominant processes involve $O_{6,8}$ operators
which the chiral structures are $(V-A)\otimes (V+A)$. The detailed
interpretation could be referred to Refs.~\cite{CKLPRD64,CGHW}
%%%%%%%%%%%%%%%%%%%%%%%%%%%%%%%%%%%%%%%%%%%%%%%%%%%%%%%%%%%%%%%%%
\begin{table}[hptb]
\caption{\label{tab:nonfa} BR (in units of $10^{-6}$), PFs and
relative phases for $B_{d}\to K^{*0} \phi$ without nonfactorization
or/and annihilation. }
\begin{ruledtabular}
\begin{tabular}{ccccccc}
 topology & $BR$& $|A_0|^2$ & $|A_{\parallel}|^2$ &
$|A_{\perp}|^2 $ &
$\phi_{\parallel}(rad)$ &  $\phi_{\perp}(rad)$ \\
\hline
     no nonfac. & $12.05$ & $0.78$ & $0.12$ & $0.10$& $2.15$ & $2.12$ \\
     no anni.  & $8.42$ & $0.83$ & $0.09$ & $0.08$& $3.30 $ & $3.32$ \\
      no both & $9.41$ & $0.92$ & $0.04$ & $0.04$& $\pi $ & $\pi$
     \\
\end{tabular}
\end{ruledtabular}
\end{table}
%%%%%%%%%%%%%%%%%%%%%%%%%%%%%%%%%%%%%%%%%%%%%%%%%%%%%%%%%%%%%%%%%

Next, we discuss the tree dominant processes $B\to \rho(\omega)
\rho(\omega)$ in which at quark level the decays are governed by
$b\to d \bar{q} q$. Since for those color-allowed decays, penguin
contributions are small, according to the analysis of
Ref.~\cite{CGHW}, it is expected that the annihilation effects are
negligible. In addition, since the nonfactorizable effects are
associated with $C_1/N_{c}$ in which $N_{c}$ is the number of color
and $C_{1}$ is roughly less than $C_{2}$ by a $N_{c}$ factor, thus,
we conjecture that the nonfactorizable contributions for
color-allowed processes are also negligible. Consequently, we
conclude that the PFs should be the same as the naive estimations,
i.e. $R_{L}\approx 1-m^{2}_{M}/m^{2}_{B}$. By using the decay
constants $f_{\rho}=f_{\omega}=200$ MeV,
$f^{T}_{\rho}=f^{T}_\omega=160$ MeV and the same taken values of
parameters for $B\to K^* \phi$, the values of $B\to \rho$ form
factors, defined by Eq.~(\ref{eq:formfactors}), in various QCD
models are given in Table~\ref{tab:brhoformfactor}.
%%%%%%%%%%%%%%%%%%%%%%%%%%%%%%%%%%%%%%%%%%%%%%%%%%%%%%%%%%%%%%%%%
\begin{table}[hptb]
\caption{ Form factors for $B\to \rho$ at $q^2=0$ in various QCD
models. }\label{tab:brhoformfactor}
\begin{ruledtabular}
\begin{tabular}{ccccc}
 Model & $V( 0)$ & $A_{0}( 0) $ & $A_{1}(0) $ & $A_{2}( 0) $
 \\ \hline
 QM \cite{MS} & $0.31$ & $0.30$ & $0.26$ & $0.24$  \\ \hline
  LCSR \cite{DA2} & $0.32$ & $0.30$ & $0.24$ & $0.22$ \\ \hline
  LFQM \cite{LF} &  $0.27$ &$0.28$ & $0.22$ &$0.20$ \\ \hline
 PQCD  & $0.26$ &$0.29$ & $0.22$ &$0.21$  \\
 \end{tabular}
\end{ruledtabular}
\end{table}
%%%%%%%%%%%%%%%%%%%%%%%%%%%%%%%%%%%%%%%%%%%%%%%%%%%%%%%%%%%%%%%%%
Again, in terms of the formulas derived by Ref.~\cite{CKLPRD66}, by
setting $\bar\Lambda=1$ GeV and by using the wave functions of
$\rho$ and $\omega$ instead of those of $\phi$ and $K^*$, the BRs,
PFs and $\phi_{\parallel(\perp)}$ of $B\to \rho(\omega)
\rho(\omega)$ are shown in Table~\ref{tab:br-po-rhorho}. The results
with conventional chosen conditions could be referred to
Ref.~\cite{Lu1}. Compare to the data displayed in
Table~\ref{tab:rhorho-data}, we find that the BR of $B_{d(u)}\to
\rho^{-}(\omega) \rho^{+}$ is consistent with the observation of
BELLE(BABAR). Although the result of $B_{u}\to \rho^{0} \rho^{+}$
doesn't fit well with current data, since the errors of data are
still large, more accumulated data are needed to further confirm. On
the other hand, in the theoretical viewpoint, the BR of $B_{u}\to
\rho^{0} \rho^{+}$ should be similar to that of $B_{u}\to \omega
\rho^{+}$. Without any anomalous effects, we still expect
$BR(B_{u}\to \rho^{0} \rho^{+})\sim BR(B_{u}\to \omega \rho^{+})$.
As for the polarizations, like our expectation, the data show that
nonfactorization and annihilation are not important in color-allowed
processes of $B\to \rho(\omega) \rho(\omega)$. We note that for
those color-suppressed decays, since the penguin effects are not
small anymore, therefore, the nonfactorizable and annihilation
effects may become important. This is the reason why we get a very
small $R_L$ in $B_{d}\to \rho^0 \rho^{0}$ decay. It is worth
mentioning that the CP asymmetry (CPA), defined by
$A_{CP}=[\Gamma(\bar B\to \bar f)-\Gamma(B\to f)]/[\Gamma(\bar B\to
\bar f)+\Gamma(B\to f)]$ with $f$ being any final state, for
$B_{d}\to \rho^{\mp} \rho^{\pm}$ has only few percent. That is, the
penguin pollution in this decay is small. Thus, we speculate that
the observed time-dependent CPA could directly indicate the bound on
the angle $\phi_{2}$ of CKM.
%%%%%%%%%%%%%%%%%%%%%%%%%%%%%%%%%%%%%%%%%%%%%%%%%%%%%%%%%%%%%%%%%
\begin{table}[hptb]
\caption{\label{tab:br-po-rhorho} BRs (in units of $10^{-6}$), PFs
and relative phases for $B\to \rho(\omega) \rho(\omega)$. }
\begin{ruledtabular}
\begin{tabular}{cccccccc}
Mode  & $BR$& $R_{L}$ & $R_{\parallel}$ & $R_{\perp} $ &
$\phi_{\parallel}(rad)$ &  $\phi_{\perp}(rad)$ &
$A_{CP}$ \\
\hline
$B^0\to \rho^{-} \rho^{+} $ & $23.06$ & $0.95$ & $0.03$ & $0.02$ & $\approx \pi$ & $\approx 0$ & $-2.96$\\
$B^0 \to \rho^0 \rho^0 $ & $0.12$ & $0.07$ & $0.43$ & $0.50$ & $3.46$ & $3.63$ & $83.21$\\
$B^0\to \rho^0 \omega $ & $0.38$ & $0.93$ & $0.04$ & $0.03$& $4.03$ & $3.93$ & $55.29$ \\
$B^0 \to \omega \omega$ & $0.35$ & $0.76$ & $0.12$ & $0.12$& $1.70$ & $1.69$ & $-92.72$\\
$B^+ \to \rho^{0} \rho^{+} $ & $11.99$ & $0.98$ & $0.01$ & $0.01$ & $\approx \pi$ & $ \approx 0$ & $\approx 0$\\
$B^+ \to \omega \rho^{+} $ & $14.78$ & $\approx 1$ & $\approx 0$ & $\approx 0$& $\approx \pi$ & $3.36$ & $-11.11$\\
%%%%%
\end{tabular}
\end{ruledtabular}
\end{table}
%%%%%%%%%%%%%%%%%%%%%%%%%%%%%%%%%%%%%%%%%%%%%%%%%%%%%%%%%%%%%%%%%
\begin{table}[hptb]
\caption{The experimental data on  BRs (in units of $10^{-6}$) and
PFs of $B\to \rho(\omega) \rho$ \cite{belle1,babar1-1,babar1-2}.
}\label{tab:rhorho-data}
\begin{ruledtabular}
\begin{tabular}{cccc}
Mode & Observation & BELLE & BABAR \\ \hline
$\rho^- \rho^{+}$&  $BR$ & $24.4\pm 2.2 ^{+3.8}_{-4.1}$ & $30\pm 4\pm 5$\\
$$&  $|A_{0}|^2$ & $0.951^{+ 0.033+0.029}_{-0.039-0.031}$ & $0.99\pm
0.03^{+0.04}_{-0.03}$\\
\hline $\rho^{0} \rho^{+} $ & BR & $31.7 \pm 7.1^{+3.8}_{-6.7}$ &
 $22.5^{+5.7} _{-5.4} \pm 5.8$\\
$$&$|A_{0}|^2$ & $0.95\pm 0.11 \pm 0.02$ & $0.97^{+0.03}_{-0.07} \pm 0.04$\\
 \hline
$\omega \rho^{+} $& BR & $--$ & $12.6^{+3.7}_{-3.3}\pm 1.6 $\\
$ $&$|A_{0}|^2$ & $--$ & $0.88^{+0.12}_{-0.15} \pm 0.03$\\
\end{tabular}
\end{ruledtabular}
\end{table}
%%%%%%%%%%%%%%%%%%%%%%%%%%%%%%%%%%%%%%%%%%%%%%%%%%%%%%%%%%%%%%%%%

Based on the previous analyses, we have learnt that by the
assumption of short-distant dominance in the $B$ decays, the
nonfactorization and annihilation are unimportant and negligible for
the tree amplitude; however, when penguin contributions are
dominant, their effects become essential on PFs. For more
comparisons with the experiments, we also calculate the results of
$B\to \rho(\omega) K^{*}$ decays. Therefore, we give the predictions
of PQCD with $\bar\Lambda=1$ GeV in Table~\ref{tab:br-po-rhoks}. In
addition, we also display the experimental data in
Table~\ref{tab:exp-rhoks}. The results by conventional PQCD could be
found in Ref.~\cite{Lu2}.
%%%%%%%%%%%%%%%%%%%%%%%%%%%%%%%%%%%%%%%%%%%%%%%%%%%%%%%%%%%%%%%%%
\begin{table}[hptb]
\caption{\label{tab:br-po-rhoks} The BRs (in units of $10^{-6}$),
PFs and relative phases for $B\to \rho (\omega) K^*$. }
\begin{ruledtabular}
\begin{tabular}{cccccccc}
Mode  & $BR$& $R_{L}$ & $R_{\parallel}$ & $R_{\perp} $ &
$\phi_{\parallel}(rad)$ &  $\phi_{\perp}(rad)$ &
$A_{CP}$\\
\hline
$B^0 \to \rho^{-} K^{*+} $ & $10.13$ & $0.60$ & $0.21$ & $0.19$& $1.60$ & $1.59$ & $-19.17$\\
$B^0 \to \rho^0 K^{*0} $ & $4.15$ & $0.70$ & $0.16$ & $0.14$ & $1.17$ & $1.17$ & $9.38$\\
$B^0 \to  \omega K^{*0} $ & $6.75$ & $0.75$ & $0.13$ & $0.12$ & $1.79$ & $1.82$ & $-7.93$\\
$B^+ \to \rho^{+} K^{*0} $ & $11.99$ & $0.78$ & $0.12$ & $0.10$ & $1.45$ & $1.46$ & $0.79$ \\
$B^+ \to \rho^{0} K^{*+}$&  $7.53$ & $0.72$ & $0.15$ & $0.13$ & $1.82$ & $1.81$ & $-19.74$\\
$B^+ \to \omega K^{*+}$&  $5.67$ & $0.61$ & $0.21$ & $0.18$ & $2.03$ & $2.06$ & $-14.31$\\
\end{tabular}
\end{ruledtabular}
\end{table}
%%%%%%%%%%%%%%%%%%%%%%%%%%%%%%%%%%%%%%%%%%%%%%%%%%%%%%%%%%%%%%%%%
%%%%%%%%%%%%%%%%%%%%%%%%%%%%%%%%%%%%%%%%%%%%%%%%%%%%%%%%%%%%%%%%%%%%%%
\begin{table}[hptb]
\caption{ The experimental data on BRs (in units of $10^{-6}$) and
PFs of $B\to \rho K^*$ \cite{babar1-1,babar4,belle4}.
}\label{tab:exp-rhoks}
\begin{ruledtabular}
\begin{tabular}{cccc}
Mode & Observation & BELLE & BABAR \\ \hline $\rho^{+} K^{*0}$ & BR
& $8.9 \pm 1.7 \pm 1.2$ & $17.0 \pm 2.9^{+2.0}_{-2.8}$\\
$$ & $|A_{0}|^2$ & $0.43 \pm 0.11^{+0.05}_{-0.02}$ & $0.79 \pm 0.08 \pm 0.04$\\
\hline $\rho^{0} K^{*+}$ & BR & $--$ & $10.6^{+3.0}_{-2.6}\pm 2.4$\\
$$ & $|A_{0}|^2$ & $--$ & $0.96^{+0.04}_{-0.15}\pm 0.04$\\
\end{tabular}
\end{ruledtabular}
\end{table}
%%%%%%%%%%%%%%%%%%%%%%%%%%%%%%%%%%%%%%%%%%%%%%%%%%%%%%%%%%%%%%%%%%%%%%
To be more clear, we summarize the main findings as follows.

  %%%%%%%%%%%%%%%%%%%
   $\bullet$ Although the decay constants $f_{\rho(K^*)}$ are larger
than $f_{\pi(K)}$, the BRs of $B\to \rho(\omega) K^*$ all are
smaller than those of $B\to \pi K$ in which the corresponding flavor
diagrams for $\pi K$ and $\rho(\omega) K^*$ in Fig.~\ref{fig:flavor}
are the same. The reason is that the factorizable contributions of
$O_{6,8}$ operators are vanished in vector-vector modes, i.e.
$\langle V_{1} V_{2}| (V-A)\otimes (V+A)|B \rangle \sim -2\langle
V_{1}|S-P|0\rangle \langle V_{2}| S+P|B\rangle=0$ due to $\langle
V_{1}|S|0\rangle =m_{V_{1}}f_{V_{1}}\epsilon_{1}\cdot P_{V_{1}}=0$
where $S(P)$ denotes the scalar(pseudoscalar) current. As a result,
the decays, which the tree amplitudes are color-allowed such as
$\rho^{\mp} K^{*\pm}$ and $\rho^{0}(\omega) K^{*\pm}$, have larger
CPAs.

 %%%%%%%%%%%%%%%%%%%%%%%%%%%
  $\bullet$ The $R_{L}$ of $B_{d}\to \rho^{-} K^{*+}$ could be as small as $60\%$.
The result could be understood as follows: since the involving tree
contributions are color-allowed, as mentioned in the decays $B\to
\rho(\omega) \rho(\omega)$, we know that the nonfactorizable effects
are negligible and transverse parts are small. Moreover, the
amplitude of penguin is opposite in sign to that of tree. Therefore,
the longitudinal part gets a large cancelation in tree and penguin
such that the $R_{L}$ is reduced. And also, the magnitude of CPA is
enhanced to be around $20\%$.

   %%%%%%%%%%%%%%%%%%%%%%%%%%%
  $\bullet$ Although the decays $B_u\to \rho^0(\omega)
K^{*+}$ possess sizable tree contributions, however besides the
diagrams Fig.~\ref{fig:flavor}(a) and (c), Fig.~\ref{fig:flavor}(b),
representing the effects of electroweak penguin mainly, also has the
contributions. And also, due to different flavor wave functions in
$\rho$ and $\omega$, respectively denoted by $(u\bar{ u}\mp
d\bar{d})/\sqrt{2}$, interestingly we find that the $R_{L}$ of
$B_u\to \rho^{0}K^{*+}$ is around $72\%$ but the $R_{L}$ of $B_u\to
\omega K^{*+}$ could be around $61\%$ which is similar to the value
of $B_{d}\to \rho^{-}K^{*+}$.

  %%%%%%%%%%%%%%%%%%%%%%%%%%%
  $\bullet$ By naive analysis, one could expect that by neglecting
the small tree contributions which are arisen from annihilation
topologies, the obtained $R_{L}$ of $B_{u}\to \rho^{+} K^{*0}$
should be similar to the value of $B_{d}\to K^{*0} \phi$. However,
the calculated results shown in the Tables~\ref{tab:br-po} and
\ref{tab:br-po-rhoks} are contrary to the expectation. The main
reason is that the sign of real part of annihilated amplitude for
$B_{u}\to \rho^{+} K^{*0}$ decay is opposite to that for $B_{d}\to
K^{*0} \phi$ decay. In other words, the annihilation is constructive
effect in $R_{L}$ of $\rho^{+} K^{*0}$ while it is destructive in
$K^{*0}\phi$. We find that the differences are ascribed to the wave
functions of mesons. In sum, the calculations of PQCD in some
physical quantities, such as PFs, strongly depend on the detailed
shapes of wave functions. Due to the sign difference in the real
part of annihilation, we predict that LPs in most $\rho (\omega)
K^*$ modes are much larger than those in $B\to K^* \phi$. We note
that the conclusion is not suitable for those tree color-allowed
processes, such as $\rho^{-} K^{*+}$ and $B_{u}\to \rho^{0}(\omega)
K^{*+}$, because according to previous discussions, the tree and/or
electroweak penguin amplitudes have significant contributions so
that the effective factors become more complicated, i.e. tree,
electroweak and annihilation all are important in these decays.

In summary, we have reanalyzed the BRs and PFs of $B\to K^* \phi$ in
the framework of PQCD. In terms of the revised conditions for the
hard scales of $B$ decays, we find the LPs of $B\to K^* \phi$ could
approach to around $60\%$ while the BRs are around $9\times
10^{-6}$. It is confirmed that the annihilation and nonfactorizable
contributions have no effects on PFs of $B\to \rho(\omega)
\rho(\omega)$ decays so that the LPs are all close to unity; and
also, we find that the BR of $B_{d(u)}\to \rho^{-}(\omega) \rho^{+}$
is consistent with the observation of BELLE(BABAR). By the
calculations, we obtain that the penguin pollution in $B_{d}\to
\rho^{\mp} \rho^{\pm}$ decays is very small so that the observed
time-dependent CPA could directly indicate the bound on the angle
$\phi_{2}$ of CKM. In addition, we also find that due to significant
tree contributions, the BR(LP) of $\rho^{-} K^{*+}$ could be around
$10.13 \times 10^{-6}(60\%)$; and due to the tree and electroweak
penguin, the BR(LP) of $\omega K^{*+}$ could be around $5.67 \times
10^{-6}(61\%)$.  \\

{\bf Acknowledgments}\\

The author would like to thank Hai-Yang Cheng, Darwin Chang,
Chao-Qiang Geng, Hsiang-Nan Li, Kingman Cheung, Chu-Khiang Chua,
Cheng-Wei Chiang and We-Fu Chang for useful discussions. This work
is supported in part by the National Science Council of R.O.C. under
Grant \#s:NSC-94-2112-M-006-009.

\end{document}